\documentclass[11pt]{article}
\usepackage[a4paper,margin=1in]{geometry}
\usepackage{amsmath,amssymb,amsthm}
\usepackage{mathpartir}
\usepackage{listings}
\usepackage{hyperref}

\theoremstyle{plain}
\newtheorem{theorem}{Theorem}[section]

\theoremstyle{definition}

\theoremstyle{remark}

\newcommand{\rta}{\rightarrow}

\newcommand{\code}[1]{\mathit{code}(#1)}
\newcommand{\sub}{\mathbf{sub}}
\newcommand{\sbf}{\mathbf{sbf}}
\newcommand{\sbt}{\mathbf{sbt}}
\newcommand{\sbftwo}{\mathbf{sbf_2}}
\newcommand{\sbttwo}{\mathbf{sbt_2}}
\newcommand{\subtwo}{\mathbf{sub_2}}
\newcommand{\thmT}{\mathbf{thmT}}
\newcommand{\num}{\mathbf{num}}

\newcommand{\seq}[1]{\langle #1 \rangle}
\newcommand{\Fun}{\mathit{Fun}}
\newcommand{\Sub}{\mathit{Sub}}

\newcommand{\Thmm}{\mathit{Thm}}
\newcommand{\Num}{\mathit{Num}}
\newcommand{\Con}{\mathit{Con}}
\newcommand{\Closed}{\mathit{Closed}}
\newcommand{\substT}{\mathit{substT}}
\newcommand{\substF}{\mathit{substF}}

\newcommand{\SN}{\mathit{SN}}

\usepackage{authblk}
\usepackage{doi}

\title{Auto formalisation of G\"odel's Second Incompleteness Theorem in Binary Recursive Arithmetic}
\author{Thierry Coquand}
\affil{Computer Science and Engineering Department,
University of Gothenburg, Sweden,
\href{mailto:coquand@chalmers.se}{coquand@chalmers.se}
}
\date{\today}

\begin{document}
\maketitle

\begin{abstract}
  We report an experiment in autoformalisation of G\"odel's second incompleteness
  theorem in Agda using Claude. The theorem is formalised for Church's Basic Recursive
  Arithmetic, following the proof outline given in Guard's 1963 lecture notes. The
  entire Agda development, comprising approximately 50,000 lines and containing no
  postulates, was produced through interaction with Claude; the author did not write
  any Agda code.

  Beyond the formalisation itself, the project provides a case study of the strengths
  and limitations of current large language models in mathematics. An initial
  autonomous attempt based on a paper of Rose failed because of a wrong Lemma;
  the resulting formal development produced by Claude established a statement
  superficially resembling G\"odel's theorem but mathematically unrelated to it. 

  The final development follows Guard's proof and required the reconstruction of several
  implicit mathematical arguments, including the role of the internal numeral-encoding
  operation and specification of substitution. The
  resulting formalisation clarifies a number of details left implicit in the original
  presentation and provides a fully machine-checked proof of Gödel's second
  incompleteness theorem for Basic Recursive Arithmetic.
\end{abstract}

\section*{Introduction}

We report an experiment in autoformalisation in Agda \cite{17Provers},
using Claude,
of G\"odel's second incompleteness theorem\footnote{From now on, we may write G\"odel I for the first incompleteness
theorem and G\"odel II for the second incompleteness theorem} for Church's \emph{basic
recursive arithmetic} (BRA), following the proof outline given by
R.~Guard in his 1963 lecture notes \cite{Guard63}. This motivation comes from
the work of J. Urban \cite{Urban26}. The goal was to see what happens
on a text which, while being precise, has several typos and gaps, and
required substantial reconstruction during formalisation.
Another contribution of this work is to document several mathematical details that had to be reconstructed
during the formalisation.

To formalise G\"odel's {\em second} incompleteness theorem  is a non-trivial formalisation project. To our knowledge, Paulson's
Isabelle formalisation \cite{Paulson14}
is the only published machine-checked proof of Gödel's second incompleteness theorem that we were able to locate.
This represents a version of this theorem not for arithmetic but for hereditary finite sets, and using a nominal
definitional extension of HOL. G\"odel himself never published Part II, announced at the end of his 1931 paper \cite{Goedel31},
where he was supposed to give a detailed proof of the second incompleteness theorem. See J. von Plato's book \cite{vonPlato20}
for further historical analysis.

The entire Agda development was produced through interaction with Claude; I did not write any Agda code myself.
The development can be found at \url{https://github.com/coquand/agda-godel-tree}. The development is done from
scratch (no library), without any tactics. Once this is done, we have enough infrastructure to encode other proofs
of G\"odel incompleteness, such as Chaitin's proof with Kolmogorov's complexity \cite{Chaitin74},
which can be found on github at the
same address and will be documented in a following paper.

\section{The original attempt}

 We started as a purely autonomous attempt using the paper of Rose \cite{Rose61}.
 We were not aware that Theorem 15 in this paper is wrong (see for instance \cite{Cook75}; this was
 correct later \cite{Rose67}).
After several days of autonomous coding, Claude was not able to prove Theorem 15,
but stated that it could find an alternative path to the G\"odel II.
We were quite impressed, until we realized that the statement claimed by Claude had very
little to do with G\"odel's result.

So it is difficult to trust the system when it is claiming to have proved a statement. It is
important that one can check independently the claimed statement.

\section{The statement}

After this attempt, we decided to follow Guard's lecture notes \cite{Guard63}. The formal system
follows Skolem's formulation of primitive recursive arithmetic \cite{Skolem23}, but with only
unary and binary function symbols. (The system is presented in details below.)

While it is a nice system for G\"odel's theorem, it would be difficult/challenging to do the formalisation
(in any proof assistant) ``by hand''.

We write $Deriv(A)$ if we can derive $A$ from these axioms and inference rules.

Following Church, we first code pairing and projection. We can then code course-of-value recursion.
While this is standard, it was non trivial for Claude, and suggestions from ChatGPT were useful.
We code directly list using pairing and never use prime decomposition like G\"odel.

We write then code for axioms and inference rules, and a function $thmT$ that enumerates the code of derived formulae
of $T$, so $thmT(O),~thmT(s(O)),\dots$ are exactly the code of derived formulae.

\section{Proof of first incompleteness Theorem}

As usual, we build by diagonalisation a formula $G$ such that $G$ is equivalent
to $thmT(x)\neq code(G)$

Intuitively, $G$ is equivalent to the fact that $G$ is not provable

It is then direct to show that $Deriv(G)$ implies $Deriv(O=s(O))$. This is a consistency free
statement of G\"odel I.

\section{Comments}

At this point, the only thing we have used of enumeration of the $thmT$ function is that
it is {\em complete}: from any proof of $Deriv(A)$ we can build $p$ such that $thmT(p) = code(A)$.

We never used that this enumeration is {\em sound}. Indeed, with the specifications written
by Guard, we cannot derive soundness.

Having an unsound version of $thmT$ there is a trivial proof of G\"odel II stated
as
$$Deriv(thmT(x_0)\neq code(0=1))\rta Deriv(0=1)$$
This is the correct statement since the variable $x_0$ is implicitly universally quantified, and so the premisse
states that there is no internal proof of $0=1$.
(Note that this is a meta-theoretic implication between derivability judgments, not a theorem of BRA.)
In the first version, $thmT$ was not supposed to be sound (since it was not stated formally in Guard's paper)
and Claude could happily use this to claim a very simple
(but trivial) proof of second incompleteness! This shows that it can (surprisingly since it can be very
clever otherwise) lack a global picture of what is going on.

The failure was not due to a missing lemma.
Claude accepted a specification of the verifier that was too weak to characterize the intended notion of provability.
It consequently produced a proof of a statement
formally resembling G\"odel's second theorem but mathematically unrelated to it.

This illustrates that current LLMs can reason correctly from incorrect specifications and may fail to detect that a formalisation
no longer captures the intended mathematical notion.

\subsection{Internal Reasoning and Surprising insights}

The proof of G\"odel II consists in {\em internalizing} this: we build a function $g$
such that $Deriv(thmT(x) = code(G) \rta thmT(g(x)) = code(O=s(O)))$

It then follows from this that $Deriv(thmT(x)\neq code(0=s(0)))$, which expressed that
the theory proves its own consistency, implies $Deriv(thmT(x)\neq code(G))$, which is equivalent
to $Deriv(G)$ by definition of $G$. Since by G\"odel I, we have $Deriv(G)$ implies $Deriv(0=1)$,
we get that $Deriv(thmT(x)\neq code(0=s(0)))$ implies $Deriv(0=1)$. This means that if the theory
proves its own consistency, it proves $0=1$ (and so is inconsistent).

For this we need to reason inside a Hilbert style system.

I always thought to use a logical system based on {\em natural deduction} was better than
working with a Hilbert-style system, with modus ponens and axioms 

\[
\begin{array}{l}
A \rta (B \rta A) \\[1ex]
\bigl(A \rta (B \rta C)\bigr)
\rta
\bigl((A \rta B) \rta (A \rta C)\bigr) \\[1ex]
(\neg A \rta \neg B) \rta (B \rta A)
\end{array}
\]

The first thing we prove as a {\em meta theorem} is the Deduction Theorem, which states that
$A_1,\dots,A_n,A\vdash B$ is equivalent to $A_1,\dots,A_n\vdash A\rta B$.

Once this is proved, we usually prove things in a natural deduction style.
For a typical example, it is easy to prove  $(A\rta (B\rta C))\rta (B\rta (A\rta C))$
using this meta theorem, while it can be difficult to derive directly this using the $S$ and $K$
axioms.

This however is a {\em meta theorem}, not something that will work here, where you have to do the
proof at the internal level.

I knew the technique of Mario Carneiro \cite{Carneiro} of {\em avoiding} this meta theorem
by systematically deriving $X\rta A$ instead of $A$, where $X$ is a variable proposition.
I communicated to Claude this method, and Claude could immediately ``understand it'' and use
it to do Hilbert style derivation. This was crucial for the feasibility of this formalisation.

After this experiment, we fully agree with the comment from metamath home page
{\em it is impressive and satisfying that we can do so much in a practical sense without stepping outside of our
Hilbert-style axiom system.}

\subsection{A subtle point in the proof}

In the proof of G\"odel I, we need to show facts such as, e.g. 
we can prove for all {\em numerals} $n$

$Deriv(id(s^n(O)) = s^{n}(O))~~~~~~~~Deriv(z(s^n(O)) = O)$

This kind of statement, proved by induction on $n$, is {\em not}
a formal statement; it mixes the meta-level 
and the object level.

The proof of G\"odel I uses in a crucial way this interplay between
use of numerals (meta theory) and internal derivation. 

In order to represent this in $T$ we need to represent this kind of ``mixed'' statements.
The problem is that our notion of coding only deals with internal statement. If we want
to internalize the proof of G\"odel I we need to {\em extend} our notion of coding, in
particular substitution.

We have an internal function $num(x)$, which plays the role of $n\mapsto s^n(O)$.

This function is introduced in Guard, but its role (mixing two levels)
is left implicit, and never discussed explicitly in Guard's paper. (This was probably discussed
during the lecture, but there is no written trace of these discussions.)
For instance, a key fact is that $num(x)$ is always closed, in the sense that it is invariant by
internal substitution. This is clear a priori since it corresponds to the fact that $S^n(O)$ is closed.
While this is intuitively clear, the complete proof is by induction on $n$, which corresponds to
an {\em internal} proof by induction on $x$ that $x$ is closed. While being crucial in the proof
of G\"odel II, this is not mentioned in \cite{Guard63}. I also needed to tell Claude to prove this fact
by internal induction on $x$.

The main Lemma for G\"odel II is then, intuitively, an internalisation of the
fact that for each unary function term $f$ and for each numeral $n$, we can prove $f~s^n(O) = s^{f(n)}(O)$.

\medskip

\section*{Conclusion of this part}

This experiment suggests that the use of LLMs may substantially change the activity of formalisation.
The language of type theory
was perfect for this attempt, where we represent combinatorial facts about syntax. We can check without problem that the
statement is correctly represented when needed. One key research question is to design formal systems where the same can be
done for parts of mathematics dealing with non-combinatorial object.

{\bf Authorship note.}
Consistent with the autoformalisation theme of this project, the rest of this text, Sections 5–11, were generated by Claude.
(I did not write a single line there.)
The mathematical content, references, and final text were reviewed by the author.


\newpage

Beyond the formalisation itself, the work
exposes a number of implicit assumptions, notational ambiguities, and
minor typographical issues in Guard's text that are worth recording
explicitly.  We also record an architectural simplification we have since
carried out: replacing the dedicated multi-variable simultaneous-substitution
machinery by nested single-variable substitution together with a
numeral-inertness lemma (the internalisation of ``a numeral is closed''),
which removes roughly ten thousand lines.  Throughout we follow Shoenfield's notation
(\emph{Mathematical Logic}, Chapter~6), with the single substitution
$\code{u}$ in place of $\ulcorner u \urcorner$.

\section{The formal system}

\subsection{Term and function syntax (Church, via Guard)}

We follow Church's basic recursive arithmetic~\cite{Church} in the
formulation of Guard~\cite{Guard63}, with Shoenfield's
notation~\cite{Shoenfield}.  Following Shoenfield {\S}6.6, the
non-logical symbols of the theory~$T$
are a constant $\mathbf{O}$, a unary symbol $\mathbf{s}$, and finitely
many additional function symbols (Church's combinators
$\mathbf{o},\mathbf{u},\mathbf{v},\mathbf{C},\mathbf{R}$), together with
the binary identity $=$.  Variables are $\mathbf{x}_0, \mathbf{x}_1,
\ldots$ in alphabetical order, with symbol number
$\SN(\mathbf{x}_i) = 2i$.

The theory~$T$ has two further function-symbol grammars:
\begin{align*}
\mathbf{f}   &::= \mathbf{s} \mid \mathbf{o} \mid \mathbf{u} \mid \mathbf{C}(\mathbf{g}, \mathbf{f_1}, \mathbf{f_2}) \quad (\text{unary, } \mathbf{f}, \mathbf{f_1}, \mathbf{f_2} \in \Fun_1), \\
\mathbf{g}   &::= \mathbf{v} \mid \mathbf{R}(\mathbf{f}, \mathbf{g_1}, \mathbf{g_2}) \quad (\text{binary, } \mathbf{g}, \mathbf{g_1}, \mathbf{g_2} \in \Fun_2).
\end{align*}
Terms~$\mathbf{t}$ are built from $\mathbf{O}$, variables
$\mathbf{x}_i$, and applications $\mathbf{f}(\mathbf{t})$ and
$\mathbf{g}(\mathbf{t_1}, \mathbf{t_2})$.

\subsection{Formulas and derivability}

Formulas are built from equations between terms by negation and
implication:
\[
\mathbf{A}, \mathbf{B} ::= \mathbf{t_1} = \mathbf{t_2} \mid \neg \mathbf{A} \mid \mathbf{A} \supset \mathbf{B}.
\]
We write $\vdash \mathbf{A}$ for ``$\mathbf{A}$ is a theorem of~$T$''
($\mathbf{A} \in \Thmm_T$ in Shoenfield's notation).  Bare $\mathbf{A}
= \mathbf{B}$ inside $\vdash$ means $\vdash \mathbf{A} = \mathbf{B}$.

\subsection{The axioms and rules, explicitly}
\label{sec:axioms}

The Hilbert calculus is exactly Guard's Definition~7~\cite{Guard63},
itself following Church's Princeton lectures~\cite{Church}.  The Agda
data type \verb|BRA3.Deriv.Deriv| has one constructor per Guard axiom and
rule and \emph{nothing else}; we list all fourteen axiom schemes and
three rules, in Guard's numbering, with the Agda constructor name in
brackets.  Here $\mathbf{f} \in \Fun_1$, $\mathbf{g} \in \Fun_2$,
$\mathbf{t},\mathbf{a},\mathbf{b},\mathbf{c},\mathbf{x},\mathbf{n}$ are
terms, $\mathbf{A},\mathbf{B},\mathbf{C}$ are formulas, and $a \in
\mathbb{N}$ is a variable index.

\medskip
\noindent\emph{Equational axioms (the defining equations of the
combinators).}
\begin{align*}
\textbf{0}\ [\mathtt{ax\_succ\_nonzero}]:\quad & \neg\,(\mathbf{s}(\mathbf{O}) = \mathbf{O}) \\
\textbf{1}\ [\mathtt{ax\_o}]:\quad & \mathbf{o}(\mathbf{t}) = \mathbf{O} \\
\textbf{2}\ [\mathtt{ax\_u}]:\quad & \mathbf{u}(\mathbf{t}) = \mathbf{t} \\
\textbf{3}\ [\mathtt{ax\_v}]:\quad & \mathbf{v}(\mathbf{a},\mathbf{b}) = \mathbf{b} \\
\textbf{8}\ [\mathtt{ax\_C}]:\quad & \mathbf{C}(\mathbf{g},\mathbf{f_1},\mathbf{f_2})(\mathbf{t}) = \mathbf{g}(\mathbf{f_1}(\mathbf{t}), \mathbf{f_2}(\mathbf{t})) \\
\textbf{9}\ [\mathtt{ax\_R\_base}]:\quad & \mathbf{R}(\mathbf{f},\mathbf{g_1},\mathbf{g_2})(\mathbf{x}, \mathbf{O}) = \mathbf{f}(\mathbf{x}) \\
\textbf{10}\ [\mathtt{ax\_R\_step}]:\quad & \mathbf{R}(\mathbf{f},\mathbf{g_1},\mathbf{g_2})(\mathbf{x}, \mathbf{s}\,\mathbf{n}) = \mathbf{g_1}\bigl(\mathbf{g_2}(\mathbf{x},\mathbf{n}),\ \mathbf{R}(\mathbf{f},\mathbf{g_1},\mathbf{g_2})(\mathbf{x},\mathbf{n})\bigr)
\end{align*}

\noindent\emph{Equality axioms.}
\begin{align*}
\textbf{4}\ [\mathtt{ax\_eqTrans}]:\quad & \mathbf{x} = \mathbf{y} \supset .\ \mathbf{x} = \mathbf{z} \supset \mathbf{y} = \mathbf{z} \\
\textbf{5}\ [\mathtt{ax\_eqCong_1}]:\quad & \mathbf{a} = \mathbf{b} \supset \mathbf{f}(\mathbf{a}) = \mathbf{f}(\mathbf{b}) \\
\textbf{6}\ [\mathtt{ax\_eqCong_L}]:\quad & \mathbf{a} = \mathbf{b} \supset \mathbf{g}(\mathbf{a},\mathbf{c}) = \mathbf{g}(\mathbf{b},\mathbf{c}) \\
\textbf{7}\ [\mathtt{ax\_eqCong_R}]:\quad & \mathbf{a} = \mathbf{b} \supset \mathbf{g}(\mathbf{c},\mathbf{a}) = \mathbf{g}(\mathbf{c},\mathbf{b})
\end{align*}

\noindent\emph{Propositional axioms.}
\begin{align*}
\textbf{11}\ [\mathtt{axK}]:\quad & \mathbf{A} \supset .\ \mathbf{B} \supset \mathbf{A} \\
\textbf{12}\ [\mathtt{axS}]:\quad & \bigl(\mathbf{A} \supset (\mathbf{B} \supset \mathbf{C})\bigr) \supset .\ (\mathbf{A} \supset \mathbf{B}) \supset .\ \mathbf{A} \supset \mathbf{C} \\
\textbf{13}\ [\mathtt{axNeg}]:\quad & (\neg\mathbf{A} \supset \neg\mathbf{B}) \supset .\ \mathbf{B} \supset \mathbf{A}
\end{align*}

\noindent\emph{Rules of inference.}
\begin{align*}
\textbf{I}\ [\mathtt{mp}]:\quad & \text{from } \mathbf{A} \supset \mathbf{B} \text{ and } \mathbf{A}, \text{ infer } \mathbf{B}. \\
\textbf{III}\ [\mathtt{ruleInst}]:\quad & \text{from } \mathbf{A}, \text{ infer } \substF(a, \mathbf{t}, \mathbf{A}) \quad (\text{Shoenfield's } \Sub). \\
\textbf{VI}\ [\mathtt{ruleIndNat}]:\quad & \text{from } \substF(a, \mathbf{O}, \mathbf{A}) \text{ and } \mathbf{A} \supset \substF(a, \mathbf{s}(\mathbf{x}_a), \mathbf{A}), \text{ infer } \mathbf{A}.
\end{align*}

Two economies of Guard's system, which we do \emph{not} exploit (we keep
all fourteen axioms as constructors for fidelity): Dana Scott showed
axiom~7 derivable from the others~\cite[Exercise after Def.~6]{Guard63},
and axiom~5 is derivable from the others excluding
axiom~7~\cite[Exercise~18]{Guard63}.  Rule~VI is stated only for
induction on $\mathbf{x}_0$ in Guard; induction on an arbitrary variable,
and the substitutivity-of-equality rule, are derivable~\cite[Exercise~19]{Guard63}.

This list is the complete trusted base: every theorem in the
development, up to and including $\mathrm{godelII}$, is built solely from
these seventeen constructors under \texttt{-{}-safe -{}-without-K -{}-exact-split},
with no postulates.

\subsection{Numerals and the encoding $\code{u}$}

For each $n \in \mathbb{N}$ the numeral is
$\mathbf{k}_n := \mathbf{s}^n(\mathbf{O})$ (Shoenfield's
$\mathbf{k}_n$).  Guard's text uses a unary internal functor
$\num \in \Fun_1$, the \emph{numeral-to-code} map: applied to a
numeral, $\num$ produces the \emph{encoding} of that numeral.  Its
defining property is the Deriv-level closure
\[
\vdash \num(\mathbf{k}_n) = \code{\mathbf{k}_n}
\quad\text{for every meta-natural $n$,}
\]
where $\code{\mathbf{k}_n}$ is the G\"odel handle of the numeral
$\mathbf{k}_n$.  This is the BRA4 lemma $\mathit{numEq}\,n$ (its Agda
form is $\vdash \num(\mathit{natCode}\,n) = \mathit{codeTerm}(\mathit{natCode}\,n)$,
where $\mathit{natCode}\,n = \mathbf{k}_n$ and
$\mathit{codeTerm} = \code{\cdot}$ on terms).

In particular $\num$ does \emph{not} act as the identity on numerals:
for instance
\[
\num(\mathbf{s}\,\mathbf{O}) = \code{\mathbf{s}\,\mathbf{O}} = \seq{\mathbf{k}_{\mathrm{tag\_ap_1}}, \seq{\mathbf{k}_{\mathrm{tag\_s}}, \mathbf{O}}} \;\neq\; \mathbf{s}\,\mathbf{O}.
\]
The functor $\num$ is precisely Guard's underline:
$\underline{\mathbf{x}} := \num(\mathbf{x})$, the code of the value of
$\mathbf{x}$.  On a numeral $\mathbf{k}_n$ this is the encoding
$\code{\mathbf{k}_n}$; on the recursion side it satisfies the
unconditional structural closures
\[
\vdash \num(\mathbf{O}) = \mathbf{O}, \qquad
\vdash \num(\mathbf{s}\,\mathbf{t}) = \seq{\mathbf{k}_{\mathrm{tag\_ap_1}}, \seq{\mathbf{k}_{\mathrm{tag\_s}}, \num(\mathbf{t})}}
\]
(the Agda lemmas $\mathbf{num\_at\_O}$, $\mathbf{num\_at\_S}$), so that
$\num$ wraps each $\mathbf{s}$ in the same nested layout that
$\code{\cdot}$ uses for $\mathbf{s}(\cdot)$, making
$\vdash \num(\mathbf{k}_n) = \code{\mathbf{k}_n}$ a straightforward
meta-induction on~$n$.

The encoding $\code{\mathbf{u}}$ of a designator $\mathbf{u}$ is
defined by induction on the formation rules, using Shoenfield's
sequence notation $\seq{a_0, a_1, \ldots, a_n}$ ({\S}6.6).  In our
concrete Agda realisation the sequence operation is Church's binary
pairing functor $\pi \in \Fun_2$, written $\seq{a, b}$, with longer
sequences nesting to the right, $\seq{a, b, \ldots} = \seq{a,\ \seq{b, \ldots}}$;
the abstract notation $\seq{\cdot}$ below leaves this implementation
choice tacit.  The encoding schema is then:
\begin{align*}
\code{\mathbf{O}}              &= \mathbf{O}, \\
\code{\mathbf{x}_k}            &= \seq{\mathbf{k}_{\mathrm{tag\_var}}, \mathbf{k}_k}, \\
\code{\mathbf{f}(\mathbf{t})}  &= \seq{\mathbf{k}_{\mathrm{tag\_ap_1}}, \code{\mathbf{f}}, \code{\mathbf{t}}}, \\
\code{\mathbf{g}(\mathbf{t_1}, \mathbf{t_2})} &= \seq{\mathbf{k}_{\mathrm{tag\_ap_2}}, \code{\mathbf{g}}, \code{\mathbf{t_1}}, \code{\mathbf{t_2}}}, \\
\code{\mathbf{t_1} = \mathbf{t_2}} &= \seq{\mathbf{k}_{\mathrm{tag\_eq}}, \code{\mathbf{t_1}}, \code{\mathbf{t_2}}}, \\
\code{\neg \mathbf{A}}         &= \seq{\mathbf{k}_{\mathrm{tag\_neg}}, \code{\mathbf{A}}}, \\
\code{\mathbf{A} \supset \mathbf{B}} &= \seq{\mathbf{k}_{\mathrm{tag\_imp}}, \code{\mathbf{A}}, \code{\mathbf{B}}},
\end{align*}
with codes for the $\Fun_1$ and $\Fun_2$ symbols defined analogously.
The tags are concrete meta-naturals (e.g.\ $\mathrm{tag\_var}=1$,
$\mathrm{tag\_ap_1}=2$, $\mathrm{tag\_eq}=10$, $\mathrm{tag\_neg}=11$,
$\mathrm{tag\_imp}=12$, etc.).

\subsection{The verifier $\thmT$ and the substitution functor $\sub$}

Guard's verifier $\thmT \in \Fun_1$ (his $\mathbf{th}$, Definition~16
of~\cite{Guard63}) is internally constructed by course-of-values
recursion so as to satisfy two structural closures:
\begin{description}
\item[Soundness-by-construction.] If $\code{\mathbf{d}}$ is the
encoding of a Hilbert derivation of~$\mathbf{A}$, then
$\vdash \thmT(\code{\mathbf{d}}) = \code{\mathbf{A}}$.

\item[Validating-decoder invariant.] On inputs that do not
encode a well-formed derivation, $\thmT$ returns the canonical truth
$\code{\mathbf{O} = \mathbf{O}}$.
\end{description}
Its construction by course-of-values recursion --- the genuinely
non-trivial step of realising such recursion inside Church's
$\{\mathbf{s},\mathbf{o},\mathbf{u},\mathbf{v},\mathbf{C},\mathbf{R}\}$
grammar --- is the subject of {\S}\ref{sec:cov}.

The substitution functor $\sub \in \Fun_2$ is Guard's Exercise~24[8]
of~\cite{Guard63}:
\[
\vdash \sub(\mathbf{z}, \code{\mathbf{A}}) = \sbf(\seq{\mathbf{k}_0, \num(\mathbf{z})}, \code{\mathbf{A}}),
\]
where $\sbf \in \Fun_2$ is the formula-level single-variable encoded
substitution.  The point of separating $\sub$ from $\sbf$ is that
$\sub$ is the function which appears in Guard's text; $\sbf$ is the
lower-level realisation.

\subsection{The diagonal lemma}

Let $\mathbf{F}$ be the seed formula
\[
\mathbf{F} \;\equiv\; \neg\,(\thmT(\mathbf{x}_1) = \mathbf{x}_0).
\]
Define
\begin{align*}
\mathbf{H} &\;\equiv\; \Sub(\code{\mathbf{F}},\, \code{\mathbf{x}_0},\, \code{\sub(\mathbf{x}_0, \mathbf{x}_0)}), \\
i &:= \Num(\code{\mathbf{H}}), \\
\mathbf{G} &\;\equiv\; \Sub(\code{\mathbf{H}},\, \code{\mathbf{x}_0},\, \code{\mathbf{k}_i}).
\end{align*}
Then $\mathbf{G}$ is closed in $\mathbf{x}_0$ with one free variable
$\mathbf{x}_1$, and satisfies the diagonal identity
\[
\vdash \sub(\mathbf{k}_i, \mathbf{k}_i) = \code{\mathbf{G}},
\]
which is the Agda lemma \texttt{Thm14F.diag\_term\_eq}.

\section{Course-of-values recursion inside Church's recursor}
\label{sec:cov}

The verifier $\thmT$ is defined by recursion on the G\"odel number of a
derivation, where the value at a node depends on the values at
\emph{all} smaller indices (the sub-derivations) --- \emph{course-of-values}
recursion.  Church's grammar provides only the primitive recursor
$\mathbf{R}$.  The reduction has a routine outline with one genuinely
delicate point.  The \emph{first step} --- express bounded \emph{iteration}
$f^{\,n}$ of a unary $f$ in terms of $\mathbf{R}$, and encode the whole
history (the table of sub-values) as a single threaded value --- is the
standard course-of-values move, and is not surprising.  What is \emph{not}
routine, and is the subject of this section, is that Church's $\mathbf{R}$
is not the textbook recursor: it returns the recursion result in a
non-standard argument slot, so even the iteration combinator $f^{\,n}$ must
be synthesised by a \emph{nested} $\mathbf{R}$ whose defining equation is
proved \emph{inside}~$T$ by induction, not by computation; and the read-off
must be shown independent of surplus recursion fuel.  Those three points
--- the slot mismatch, the internally-proved iteration combinator, and
fuel-stability --- are the genuine content.

\subsection{Church's recursor is not the textbook one}

The textbook primitive recursor satisfies $\rho(\mathbf{x},\mathbf{O}) =
g(\mathbf{x})$ and $\rho(\mathbf{x},\mathbf{s}\,\mathbf{n}) =
h(\mathbf{n}, \rho(\mathbf{x},\mathbf{n}))$, feeding the recursion result
to $h$ in a fixed argument slot.  Church's $\mathbf{R}$ (axioms~9,~10 of
{\S}\ref{sec:axioms}) instead satisfies
\[
\mathbf{R}(\mathbf{f},\mathbf{g_1},\mathbf{g_2})(\mathbf{x},\mathbf{s}\,\mathbf{n}) = \mathbf{g_1}\bigl(\mathbf{g_2}(\mathbf{x},\mathbf{n}),\ \mathbf{R}(\mathbf{f},\mathbf{g_1},\mathbf{g_2})(\mathbf{x},\mathbf{n})\bigr),
\]
so the recursion result enters $\mathbf{g_1}$ as its \emph{second}
argument, while $\mathbf{g_2}(\mathbf{x},\mathbf{n})$ supplies an
auxiliary first argument.  This shape is convenient for some recursions
and awkward for others; in particular it does \emph{not} directly give
``iterate a unary $\mathbf{f}$,'' which is what course-of-values needs.

\subsection{The history-as-value encoding}

Following Church's pairing functors $J,K,L$ (Definition~6
of~\cite{Guard63}; the realisation uses Church's $\pi$, written
$\seq{a,b}$, with projections $\mathit{Fst},\mathit{Snd}$), the history of
the values at nodes $0,\ldots,n$ is encoded as the right-nested list
\[
\mathit{table}_n \;=\; \seq{v_n, v_{n-1}, \ldots, v_0} \;=\; \seq{\,v_n,\ \seq{v_{n-1},\ \cdots \seq{v_0,\ \mathbf{O}}}\,},
\]
and the recursion carries a \emph{state}
\[
\mathit{state}_n \;=\; \seq{\,\mathbf{k}_n,\ p,\ \mathit{table}_n\,}.
\]
This threads three things: the counter $\mathbf{k}_n$ (the numeral of the
current node $n$); an immutable parameter $p$, recoverable as
$\mathit{Fst}(\mathit{Snd}(\mathit{state}))$ at every step --- this is how
the step function reads a fixed parameter \emph{without} it being a BRA
argument; and the full table $\mathit{table}_n$.  (The parameter $p$ is
what the Agda calls the \emph{spec}: for $\sbt/\sbf$ it carries the
substitution specification $\seq{k,S}$, and for $\thmT$ it is an unused
dummy $\mathbf{O}$; we write it $p$.)  At each step the new value
$v_{n+1} = \mathit{stepFun}(\mathit{Fst}\,\mathit{state},\,
\mathit{Snd}\,\mathit{state})$ is consed onto the table; the final answer
is read off the head,
\[
\thmT = \mathit{Post}\;\mathit{readOff}\;(\mathit{cov\_spec}\ \mathit{baseFun}\ \mathit{stepFun}).
\]
This is the Agda module \verb|BRA4.CoVSpec| (the ``spec'' in the name is
this parameter $p$), built on \verb|BRA3.CourseOfValues|.

\subsection{The nested-\texorpdfstring{$\mathbf{R}$}{R} trick}

The crux is that the state recursion needs to \emph{apply a unary
function to the previous state} --- i.e.\ an iteration combinator
$\mathit{iter}\ f$ with $\mathit{iter}\ f\,(\mathbf{x},\mathbf{k}) =
f^{\mathbf{k}}(\mathbf{x})$.  But Church's $\mathbf{R}$ delivers the
recursion result into $\mathbf{g_1}$'s \emph{second} slot, whereas the
naive lift $\mathbf{R}\,f\,\mathbf{v}\,\mathbf{v}$ can read it only via
the \emph{first}.  Bridging the two requires a binary $\mathbf{g_1}$ with
$\mathbf{g_1}(\mathbf{a},\mathbf{b}) = f(\mathbf{b})$ --- a function that
ignores its first argument and applies $f$ to its second.  This is itself
built as \emph{another} nested $\mathbf{R}$,
\[
\mathit{iter\_step\_fun}\,f \;=\; \mathbf{R}\,\bigl(\mathit{compose_1}\,f\,\mathbf{o}\bigr)\,\bigl(\mathbf{R}\,(\mathit{compose_1}\,f\,\mathbf{s})\,\mathbf{v}\,\mathbf{v}\bigr)\,\mathbf{v},
\]
and the defining equation $\mathbf{g_1}(\mathbf{a},\mathbf{b}) =
f(\mathbf{b})$ is then \emph{proved inside $T$ by induction}
(\texttt{ruleIndNat}, Rule~VI), not by computation.  This use of the
object-level induction rule to establish a meta-level recursion scheme is
exactly what makes the step ``non-trivial to do for Church's system'':
the iteration combinator is not primitive, it is a derived object whose
correctness is an internal theorem.

\subsection{Fuel and stability}

Church's $\mathbf{R}$ recurses down to a \emph{given} numeral, so
$\mathit{cov\_spec}$ must be run with an explicit \emph{fuel} argument: a
numeral bounding how far the history is built.  Evaluating $\thmT(y)$ needs
the recursion run to at least node $y$ --- the value for $y$ sits at depth
$y$ in the history --- and since no tight bound on the required depth is
known in advance, one runs with \emph{surplus} fuel and must show the answer
does not depend on it.

That is the content of the \emph{stability} lemmas
(\verb|BRA4.StabilityNatFuel|, \verb|BRA4.Stability|): once the counter has
reached the target node $n$, building further history leaves the value
already recorded at $n$ unchanged,
\[
\mathit{readOff}(\mathit{state}_{n'}) \;=\; \mathit{readOff}(\mathit{state}_n)
\qquad\text{for all } n' \ge n .
\]
Equivalently, the read-off at a node is insensitive to surplus fuel above
it.  Together with the per-projection preservation lemmas, this is what lets
the finitely-many per-tag closure equations of $\thmT$
($\thmT\_\mathrm{at}\_\mathrm{ax_0}, \ldots, \thmT\_\mathrm{at}\_\mathrm{mp},
\thmT\_\mathrm{at}\_\mathrm{sb}$) be stated and discharged for an arbitrary
sufficiently-large fuel, rather than a hand-computed one.

In sum: course-of-values recursion is \emph{representable} in BRA, but not
for free.  Its representation requires (i)~a history-carrying state threaded
through Church's non-standard $\mathbf{R}$, (ii)~the iteration combinator of
the nested-$\mathbf{R}$ trick, whose correctness is an internal induction,
and (iii)~the fuel-stability lemmas above.  This is the foundation on which
$\thmT$, $\sbt$, $\sbf$, $\num$, and $\mathbf{Eval}$ (of the Berry route)
all rest.

\section{The main theorem}

\begin{theorem}[G\"odel II for BRA, formalised]
\label{thm:godelII}
Let
\[
\Con_T \;\equiv\; \neg\,(\thmT(\mathbf{x}_0) = \code{\mathbf{O} = \mathbf{s}\,\mathbf{O}}).
\]
This is an \emph{open} formula, with the single free variable
$\mathbf{x}_0$.  BRA has no object quantifiers ({\S}\ref{sec:axioms}): a
theorem is an open formula, read as \emph{implicitly universally
quantified} over its free variables --- $\vdash\mathbf{A}(\mathbf{x}_0)$
delivers, by Rule~III, every closed instance $\vdash\mathbf{A}(\mathbf{t})$.
So $\Con_T$, as that single open formula, \emph{is} the internal
consistency statement: \emph{for every} term $\mathbf{x}_0$, $\mathbf{x}_0$
does not code a derivation of $\mathbf{O} = \mathbf{s}\,\mathbf{O}$.  Then
\[
\mathrm{godelII} \;:\quad {\vdash}\,\Con_T \;\Longrightarrow\; {\vdash}\,(\mathbf{O} = \mathbf{s}\,\mathbf{O}).
\]
\end{theorem}

\noindent
The arrow ``$\Longrightarrow$'' is a \emph{meta-level} implication, not an
object connective.  What is proved is a \emph{meta-statement} ---
equivalently, a total function transforming any closed derivation of
$\Con_T$ into a closed derivation of $\mathbf{O} = \mathbf{s}\,\mathbf{O}$
--- and \emph{not} a single closed Hilbert-style derivation of the object
implication ``$\Con_T \supset (\mathbf{O} = \mathbf{s}\,\mathbf{O})$''.
This is exactly what the Agda type records: $\mathrm{godelII}$ is an arrow
of the meta-language Agda, with \texttt{Deriv} the (object) derivability
relation and \verb|->| Agda's function arrow:
\begin{lstlisting}
godelII : Deriv ConSchema -> Deriv P_false
  where ConSchema = neg (eqF (ap1 thmT (var 0)) codeFalse)
        P_false   = atomic (eqn O (ap1 s O))
\end{lstlisting}
and is shipped in \verb|BRA4/Thm/Thm14GodelII.agda| (241 LoC, fresh
chain to the theorem~$7.2$\,s), with no postulates, no holes, and the
global options \verb|--safe --without-K --exact-split|.  The full BRA4
development has zero \verb|postulate| declarations anywhere in its
$\sim 50\,000$ lines.

\section{Structure of the proof}

The proof follows Guard's five-step decomposition (\emph{guard15.pdf}
p.~17), with Step~5 expressed as an explicit composition of two encoded
applications of modus ponens.  Throughout, $\mathbf{x}$ ranges over
$T$-terms at the meta level, and is \emph{universally} quantified at
the meta-level (so $\vdash \mathbf{A}(\mathbf{x})$ below stands for
``\,$\vdash \mathbf{A}$ as an open formula with free variable
$\mathbf{x}$\,'' --- giving every closed instance via Rule~III).

\paragraph{Step 1.}  By the internal Theorem~13 for unary functionals,
\[
\vdash \;\;(\thmT(\mathbf{x}) = \mathbf{k}_j) \;\supset\; (\thmT(D_\thmT(\mathbf{x})) = \code{\thmT(\num(\mathbf{x})) = \num(\mathbf{k}_j)}),
\]
where $D_\thmT \in \Fun_1$ is the Hilbert-internal handle for the
verifier $\thmT$ supplied by Theorem~12, and $j$ is the Goedel-handle
meta-natural of~$\mathbf{G}$.

\paragraph{Step 2.}  By the internal Theorem~13 for the binary
$\sub$:
\[
\vdash \;\;\thmT(D_\sub(\mathbf{k}_i, \mathbf{k}_i)) = \code{\sub(\mathbf{k}_i, \mathbf{k}_i) = \num(\mathbf{k}_j)}.
\]

\paragraph{Step 3.}  Imp-encoded transitivity of equality chains
Steps~1 and~2 through the common middle term $\num(\mathbf{k}_j)$:
\[
\vdash \;\;(\thmT(\mathbf{x}) = \mathbf{k}_j) \;\supset\; (\thmT(g(\mathbf{x})) = \code{\thmT(\num(\mathbf{x})) = \sub(\mathbf{k}_i, \mathbf{k}_i)}).
\]

\paragraph{Step 4.}  The term
\[
K_{\mathrm{part}}(\mathbf{x}) \;:=\; \seq{\mathbf{k}_{\mathrm{tag\_sb}}, \seq{\mathbf{k}_1, \num(\mathbf{x})}, \mathbf{x}}
\]
is the encoded single-variable substitution that, under the diagonal
identity $\sub(\mathbf{k}_i, \mathbf{k}_i) = \code{\mathbf{G}}$,
satisfies
\[
\vdash \;\;(\thmT(\mathbf{x}) = \mathbf{k}_j) \;\supset\; (\thmT(K_{\mathrm{part}}(\mathbf{x})) = \code{\neg\,(\thmT(\num(\mathbf{x})) = \sub(\mathbf{k}_i, \mathbf{k}_i))}).
\]

\paragraph{Step 5a.}  Let $t' := \code{\mathbf{axExFalso}\,\mathbf{A}\,\bot}$
be the encoded derivation of Guard's schema $\mathbf{A} \supset
(\neg \mathbf{A} \supset \bot)$ with $\mathbf{A} \equiv (\mathbf{x}_0
= \mathbf{x}_1)$ and $\bot \equiv (\mathbf{O} = \mathbf{s}\,\mathbf{O})$.
Set $h(\mathbf{x})$ to be \emph{two nested single-variable substitution
wraps} of $t'$ (var~$0$ outermost, var~$1$ innermost):
\[
h(\mathbf{x}) \;:=\; \seq{\mathbf{k}_{\mathrm{tag\_sb}},\, \seq{\mathbf{k}_0, \code{\thmT(\num(\mathbf{x}))}},\,
   \seq{\mathbf{k}_{\mathrm{tag\_sb}},\, \seq{\mathbf{k}_1, \code{\sub(\mathbf{k}_i, \mathbf{k}_i)}},\, t'}}.
\]
$\thmT$ decodes the outer wrap by its single $\mathbf{sb}$-clause into
$\sbf(\seq{\mathbf{k}_0, \code{\thmT(\num\,\mathbf{x})}},\, \cdot)$ applied to
the inner wrap, which decodes likewise into
$\sbf(\seq{\mathbf{k}_1, \code{\sub(\mathbf{k}_i,\mathbf{k}_i)}},\, \thmT(t'))$.
Both substituents are $\num$-based (var-free) codes, hence inert under the
outer re-scan ({\S}\ref{sec:sb2}); writing
$E := \thmT(\num(\mathbf{x})) = \sub(\mathbf{k}_i, \mathbf{k}_i)$,
\[
\vdash \;\;\thmT(h(\mathbf{x})) = \code{E \;\supset\; \neg E \;\supset\; (\mathbf{O} = \mathbf{s}\,\mathbf{O})}.
\]

\paragraph{Step 5b.}  Two applications of the Carneiro-lifted encoded
modus ponens $\mathbf{imp\_encoded\_mp}$ compose Steps~3, 4, and~5a to
produce
\[
\vdash \;\;(\thmT(\mathbf{x}) = \mathbf{k}_j) \;\supset\; (\thmT(\mathit{bigterm}(\mathbf{x})) = \code{\mathbf{O} = \mathbf{s}\,\mathbf{O}}),
\]
where
\[
\mathit{bigterm}(\mathbf{x}) \;:=\; \seq{\mathbf{k}_{\mathrm{tag\_mp}}, \seq{\mathbf{k}_{\mathrm{tag\_mp}}, h(\mathbf{x}), g(\mathbf{x})}, K_{\mathrm{part}}(\mathbf{x})}.
\]

\paragraph{Final assembly.}  Instantiate the universal Step~5 at
$\mathbf{x} = \mathbf{x}_1$.  From $\vdash \Con_T$ by Rule~III applied
to $\mathit{bigterm}(\mathbf{x}_1)$ we obtain
\[
\vdash \;\neg\,(\thmT(\mathit{bigterm}(\mathbf{x}_1)) = \code{\mathbf{O} = \mathbf{s}\,\mathbf{O}}).
\]
The classical contrapositive $\mathbf{axContrapos}$ then yields
\[
\vdash \;\neg\,(\thmT(\mathbf{x}_1) = \mathbf{k}_j).
\]
The predicate-Leibniz rule $\mathbf{substF\_cong}$ applied at the
formula
\[
\Phi \;\equiv\; \neg\,(\thmT(\mathbf{x}_1) = \mathbf{x}_2)
\]
with the equation $\vdash \sub(\mathbf{k}_i, \mathbf{k}_i) = \mathbf{k}_j$
converts this to $\vdash \mathbf{G}$.  An inlined re-derivation of
Theorem~11 (G\"odel~I) for the sub-form of $\mathbf{G}$ then yields
$\vdash (\mathbf{O} = \mathbf{s}\,\mathbf{O})$.

\section{Implicit facts in Guard's text, made explicit}

The formalisation forced a number of small but non-trivial corrections,
disambiguations, or additions to Guard 1963.

\subsection{Underline notation: the internal $\num$ versus the meta-encoding $\code{\cdot}$}

Guard's text uses an underline notation $\underline{\mathbf{x}}$ whose
meaning shifts between (i)~the internal $T$-term $\num(\mathbf{x})$
and (ii)~the meta-level G\"odel handle $\code{\mathbf{x}}$.  These are
genuinely different objects:
\begin{itemize}
\item $\num(\mathbf{x})$ is a $T$-term --- the unary functor $\num \in
\Fun_1$ applied to $\mathbf{x}$.  It \emph{computes}, inside $T$, the
encoding of the value of $\mathbf{x}$.
\item $\code{\mathbf{x}}$ is the meta-level encoding of the
\emph{syntactic} object $\mathbf{x}$ (the G\"odel handle), a closed
term computed outside $T$.
\end{itemize}
They are connected only by the Deriv-level closure on closed numerals:
\[
\vdash \num(\mathbf{k}_n) = \code{\mathbf{k}_n}
\quad\text{(the lemma $\mathit{numEq}\,n$),}
\]
and they differ for non-numeral $\mathbf{x}$ (e.g.\ for a variable
$\mathbf{x}_0$, the term $\num(\mathbf{x}_0)$ is open whereas
$\code{\mathbf{x}_0} = \seq{\mathbf{k}_{\mathrm{tag\_var}}, \mathbf{k}_0}$
is a fixed closed handle).  Note that \emph{neither} of these equals
$\mathbf{x}$ itself: in particular $\num(\mathbf{s}\,\mathbf{O}) =
\code{\mathbf{s}\,\mathbf{O}} \neq \mathbf{s}\,\mathbf{O}$, so the
underline is never the identity.

In the Agda development the file \verb|Thm14F.agda| introduces the
\emph{slot abbreviation} $\mathrm{code}\,\mathbf{t} := \num(\mathbf{t})$
(its $\mathrm{code}$ is literally $\num$), used in the encoded
positions where Guard writes $\underline{\mathbf{x}}$.  This
abbreviation coincides with the meta-encoding $\code{\cdot}$ only on
closed numerals --- precisely via $\mathit{numEq}$ --- and the
formalisation makes every such bridge explicit
(e.g.\ $\vdash \num(\mathbf{k}_i) = \code{\mathbf{k}_i}$ at the
diagonal numeral $i$).

\subsection{The two G's of the diagonal lemma}

There are two distinct formulas, both called ``$\mathbf{G}$'' in
informal treatments of the diagonal lemma:
\begin{align*}
\mathbf{G}_{\sub}  &:\equiv\; \neg\,(\thmT(\mathbf{x}_1) = \sub(\mathbf{k}_i, \mathbf{k}_i)), \\
\mathbf{G}_{\sbf}  &:\equiv\; \neg\,(\thmT(\mathbf{x}_1) = \sbf(\seq{\mathbf{k}_0, \num(\mathbf{k}_i)}, \mathbf{k}_i)).
\end{align*}
These are equal in~$T$ only up to the closure $\mathbf{sub\_eq}$,
which is a term-level $T$-derivable equation, \emph{not} a
Formula-syntactic equality.  Consequently, a proof of G\"odel~I phrased
for $\mathbf{G}_{\sbf}$ (as in our \verb|BRA4.GoedelI|) does not
literally apply to $\mathbf{G}_{\sub}$.  The formalisation bridges this
gap either by inlining a copy of G\"odel~I for $\mathbf{G}_{\sub}$ (the
route taken in \verb|Thm14GodelII|), or by invoking the
\emph{predicate-Leibniz} rule $\mathbf{substF\_cong}$ to transfer
derivability across the term-equation $\sub(\mathbf{k}_i, \mathbf{k}_i)
= \mathbf{k}_j$.  Guard's text treats the two $\mathbf{G}$'s as
interchangeable without comment.

\subsection{Modus-ponens slot order}

{\sloppy Guard writes \emph{mp}(antecedent, implication) in
his encoded-derivation tables, while our
$\seq{\mathbf{k}_{\mathrm{tag\_mp}}, \mathrm{impl}, \mathrm{ant}}$ uses
the opposite order.\par}  Guard's expression for
$\mathit{bigterm}(\mathbf{x})$,
\[
4J\bigl[\,4J(J(\num\,\mathbf{x},1),\, \mathbf{x})+1,\ \ 4J(g(\mathbf{x}),\, h(\mathbf{x}))+2\,\bigr]+2,
\]
thus hides which sub-slot is which, and a careful sign-by-sign
translation is needed.

\subsection{Closedness side conditions for Rule III}

Rule III ($\substT, \substF$) only reduces $\substT(\mathbf{x}_a,
\mathbf{t}, \mathbf{t}')$ when $\mathbf{x}_a$ and $\mathbf{t}'$ match
by definitional pattern.  For closed sub-terms (numerals, the encoded
false formula $\code{\mathbf{O} = \mathbf{s}\,\mathbf{O}}$, the
diagonal numeral $\mathbf{k}_i$) the formalisation requires explicit
$\Closed$ witnesses and bridges via \texttt{closedAt}.  Guard's text
uses ``\,$\mathbf{k}_i$ is a numeral so its substitution is
trivial\,''.  The Agda development must construct, e.g.\
\[
\mathit{closed\_codeFalse} \;:\; \Closed\,\code{\mathbf{O} = \mathbf{s}\,\mathbf{O}}
\]
as a nested $\seq{\cdot}$/numeral proof tree.

\subsection{The validating-decoder invariant on $\thmT$}

Guard's verifier $\thmT$ is described informally as ``returns the code
of the proved formula''.  But for inputs that are \emph{not} the
encoding of a derivation, $\thmT$ must still be defined and must still
satisfy the closure equations used in the proof (e.g.\
$\vdash \thmT(\code{\mathrm{mp}(p,q)}) = \cdots$).  We adopt the
convention
\[
\vdash \thmT(\mathit{garbage}) = \code{\mathbf{O} = \mathbf{O}},
\]
(the canonical trivial truth), which makes the closure equations
$\mathbf{thmT\_at\_mp}$, $\mathbf{thmT\_at\_sb}$, etc.\ universal in
their raw $\seq{\cdot}$-encoded inputs.  This invariant is implicit in Guard
but essential to make the closures usable.

\subsection{The encoded $t'$ schema}

Guard writes ``\,$t'$ is the encoded derivation of $\mathbf{x}_0 =
\mathbf{x}_1 \supset \mathbf{x}_0 \neq \mathbf{x}_1 \supset \mathbf{O}
= \mathbf{s}\,\mathbf{O}$\,''.  The formalisation exhibits this
concretely as
\[
t' \;=\; \code{\mathbf{axExFalso}\,(\mathbf{x}_0 = \mathbf{x}_1)\,(\mathbf{O} = \mathbf{s}\,\mathbf{O})}
\]
with $\mathbf{axExFalso}$ built from $\mathbf{axK}, \mathbf{axS},
\mathbf{axNeg}$, and a completeness handle
$\mathbf{thmT\_complete\_rec}$ proving the verifier's expected output.
Guard glosses both pieces.

\subsection{Carneiro lift for hypothetical derivations}
\label{sec:carneiro}

Guard frequently asserts implications of the form ``\,under the
hypothesis $\mathbf{P}$ we have $\mathbf{A}$ and $\mathbf{B}$, hence
$\mathbf{C}$\,''.  In a pure Hilbert calculus there is no
implication-introduction rule, and our \verb|Deriv| is
\emph{context-free} ($\verb|Deriv| : \mathit{Formula} \to \mathit{Set}$,
no hypothesis context), so the only way to express ``under hypothesis
$\mathbf{P}$'' is the object formula $\mathbf{P} \supset (\cdots)$, and
the only way to \emph{compose} such conditionals is to lift every
ordinary combinator to its $\mathbf{P} \supset (\cdots)$ form via
Carneiro's $\mathbf{axS}/\mathbf{axK}$ technique~\cite{Carneiro}.  The
formalisation extracts this into a small toolkit in
\verb|BRA4.Thm12.ImpHelpers|:
\[
\mathbf{impLift},\ \mathbf{impMp},\ \mathbf{impCong_1},\ \mathbf{impCong_L},\ \mathbf{impCong_R},\ \mathbf{impRuleSym},\ \mathbf{impEqTrans},
\]
together with the encoded modus ponens $\mathbf{imp\_encoded\_mp}$
(\verb|BRA4.Thm12.EncodedMp|).

\paragraph{Is it essential?}  Relative to the context-free
\verb|Deriv| representation, yes: the theorems of the Theorem~14
argument are intrinsically conditional --- Step~5 is literally
$\mathrm{th}(\mathbf{x}) = j \supset \mathrm{th}(\mathit{bigterm}(\mathbf{x}))
= \code{\mathbf{O}=\mathbf{s}\,\mathbf{O}}$ --- and this conditional is
exactly what is combined with $\Con_T$ to produce $\mathbf{G}$, so it
cannot be restructured away.  It is \emph{not} essential in the absolute
sense: a deduction theorem for BRA is admissible (\cite[Exercise~19]{Guard63}),
so one could instead carry a hypothesis context in \verb|Deriv| and prove
the deduction theorem once.  The Carneiro lift is the \emph{pointwise}
deduction theorem --- lift the combinators actually used, rather than
prove the general metatheorem --- and given our context-free judgement it
is the lightweight, effectively forced choice.

\paragraph{A concrete example where it helped.}  The step in
$\thmT$'s Step~4 (\verb|BRA4.Thm.Thm14Step4|) that \emph{uses the
G\"odel hypothesis itself}.  Under $\mathbf{P} = (\mathrm{th}(\mathbf{x})
= j)$ one must rewrite $\sbf(\mathit{cSpec}, \mathrm{th}(\mathbf{x}))$
into $\sbf(\mathit{cSpec}, j)$ --- i.e.\ apply the assumption
``$\mathrm{th}(\mathbf{x}) = j$'' \emph{inside} a congruence, without
discharging it:
\begin{lstlisting}
hyp_imp : Deriv (imp P (eqF (ap1 thmT x) j))
hyp_imp = impRefl P             -- assuming P, P holds: i.e. assuming th(x)=j, th(x)=j

step_B_imp : Deriv (imp P (eqF (ap2 sbf cSpec (ap1 thmT x)) (ap2 sbf cSpec j)))
step_B_imp = impCongR {P} sbf (ap1 thmT x) j cSpec hyp_imp
\end{lstlisting}
Here $\mathbf{impRefl}\,\mathbf{P}$ extracts the hypothesis as
$\mathbf{P} \supset \mathbf{P}$ --- and $\mathbf{P}$ \emph{is} the
equation $\mathrm{th}(\mathbf{x}) = j$ --- while $\mathbf{impCong_R}$
threads it through the congruence of $\sbf$, yielding
$\mathbf{P} \supset (\sbf(\mathit{cSpec}, \mathrm{th}(\mathbf{x})) =
\sbf(\mathit{cSpec}, j))$.  Unfolded, this is just $\mathbf{ax\_eqCong_R}$
wrapped with $\mathbf{axS}/\mathbf{axK}$; the toolkit makes it one line.

The same proof shows the lift is doing genuine work of two distinct
kinds at once.  The surrounding steps A, C, D of Step~4 are
\emph{unconditional} facts (the closure $\thmT\_\mathrm{at}\_\mathrm{sb}$,
the diagonal bridge $\mathit{codeFormulaG\_eq\_j}$, the per-shape
$\sbf/\sbt$ evaluations); each is carried past $\mathbf{P}$ by
$\mathbf{impLift}$ (a bare $\mathbf{axK}$ wrap).  Only step~B
\emph{consumes} the hypothesis ($\mathbf{impRefl}\,\mathbf{P}$ +
$\mathbf{impCong_R}$).  And $\mathbf{impEqTrans}$ chains all four equalities
under the single $\mathbf{P}$ to conclude
$\mathbf{P} \supset (\mathrm{th}(K_{\mathrm{part}}(\mathbf{x})) =
\code{\neg(\mathrm{th}(\underline{\mathbf{x}}) = \sub(\mathbf{k}_i,\mathbf{k}_i))})$.
Carrying unconditional facts past $\mathbf{P}$ and applying $\mathbf{P}$
where needed is exactly the bookkeeping a deduction theorem would
otherwise perform; the Carneiro lift does it pointwise.  Its headline
consumer is $\mathbf{imp\_encoded\_mp}$: the two encoded modus-ponens
steps of Step~5b are composed entirely under $P_x$ via
$\mathbf{imp\_thmT\_at\_mp} + \mathbf{impEqTrans}$.  Guard never names the
principle.

\subsection{Predicate-Leibniz substitutivity}

Guard's transition from
``\,$\neg\,(\thmT(\mathbf{x}_1) = \mathbf{k}_j)$\,''
to
``\,$\neg\,(\thmT(\mathbf{x}_1) = \sub(\mathbf{k}_i, \mathbf{k}_i))$\,''
is stated as an obvious rewrite.  In a Hilbert calculus this requires
the \emph{predicate-Leibniz} principle
\[
\vdash \mathbf{a} = \mathbf{b} \;\Longrightarrow\; \vdash \Sub(\code{\Phi}, \code{\mathbf{x}_k}, \code{\mathbf{a}}) \supset \Sub(\code{\Phi}, \code{\mathbf{x}_k}, \code{\mathbf{b}}),
\]
which is itself a small but non-trivial Hilbert-meta-theorem (proved by
induction on $\Phi$, using $\mathbf{axContrapos}$ on the negative
slot; see \verb|BRA3.Substitutivity|).  Guard does not mention it.

\subsection{Three small typographical issues}

\begin{itemize}
\item Guard writes ``$j$'' in places where his own underline convention
requires $\underline{j}$ (= $\num\,\mathbf{k}_j = \code{\mathbf{k}_j}$).
The formalisation makes the distinction explicit at every occurrence.

\item Definition~12 conflates the underline notation with the
numeral.  The formalisation requires the precise statement that
$\vdash \num(\mathbf{k}_n) = \code{\mathbf{k}_n}$ \emph{for every
meta-natural~$n$} --- i.e.\ $\num$ maps a numeral to its
\emph{encoding}, not to itself --- instantiated through the lemma
$\mathit{numEq}$.  In particular $\num(\mathbf{s}\,\mathbf{O}) =
\code{\mathbf{s}\,\mathbf{O}} \neq \mathbf{s}\,\mathbf{O}$.  The text
leaves this implicit and the underline notation makes it easy to
mistake $\num(\mathbf{x})$ for $\mathbf{x}$ itself.

\item In the description of $h(\mathbf{x})$, Guard does not commit to a
representation of the substitution.  The formalisation realises the
two-variable effect as two \emph{nested single-variable} wraps
$\seq{\mathbf{k}_{\mathrm{tag\_sb}}, \seq{\mathbf{k}_0, S_0},
\seq{\mathbf{k}_{\mathrm{tag\_sb}}, \seq{\mathbf{k}_1, S_1}, t}}$ and uses
the numeral-inertness lemma ({\S}\ref{sec:sb2}) to pass the inner
substituent $S_1$ through the outer re-scan.
\end{itemize}

\section{Single-variable substitution suffices: numeral-inertness}
\label{sec:sb2}

The verifier's substitution clause is the \emph{single-variable} encoded
substitution $\sbf/\sbt$ (cf.\ Guard's $\mathbf{sb}$, Def.~16).  Both the
compound-functor congruences feeding Theorem~12 and the two-variable wrap of
Step~5a require, on the face of it, a \emph{simultaneous} substitution of
several variables at once.  We obtain all of them from \emph{nested
single-$\mathbf{sb}$ wraps}, with no dedicated multi-variable functor.  The
one fact that makes this work is a numeral-inertness lemma; we record it
here because it is exactly the place Guard's text leaves the argument
implicit, and because the alternative (a genuine simultaneous-substitution
primitive) carries a different proof-theoretic commitment relevant to one of
Guard's open problems.

\paragraph{The apparent obstruction.}  A nested wrap substitutes the
variables one at a time, so an outer pass re-scans a substituent already
plopped by an inner pass.  In Theorem~12's compound cases the substituents
are $\num\,X$ for the \emph{object} variable $X$, and for open $X$ (e.g.\
$X=\mathbf{x}_1$) the term $\num(\mathbf{x}_1)$ does not reduce:
$\num = \mathbf{C}\,\mathit{numAux}\,\mathbf{o}\,\mathbf{u}$ with
$\mathit{numAux}=\mathbf{R}\,\mathbf{o}\,\mathit{numStep}\,\mathbf{v}$, whose
$\mathbf{R}$-recursion fires only when its second argument is syntactically
$\mathbf{O}$ or $\mathbf{s}(\cdot)$; a variable is neither, so
$\num(\mathbf{x}_1)$ stays a stuck open application.  Since $\sbt$ dispatches
on $\mathit{get\_tag}(\mathit{input})$ and that tag is an \emph{open}
expression on the stuck term, naive evaluation of
$\sbt(\seq{\mathbf{k}_k, S}, \num(\mathbf{x}_1))$ never settles.  Read
\emph{computationally}, the re-scan stalls --- which is why a simultaneous
plop (no re-scan) looks unavoidable.

\paragraph{The numeral-inertness lemma.}  The obstruction conflates
computation with \emph{derivability}.  We never need $\sbt$ to \emph{compute}
on a stuck $\num(\mathbf{x}_1)$; we need the \emph{equation}
\[
\vdash\ \sbt(\seq{\mathbf{k}_k, S},\, \num\,X) \;=\; \num\,X
\qquad\text{for an \emph{arbitrary} object term } X
\]
(the Agda lemma $\mathbf{sbt\_num\_inert}$, \verb|BRA4.NumInert|), and this is
a \emph{theorem} of the object system, proved by \emph{internal} induction on
$X$ (Rule~III / $\mathbf{ruleIndNat}$), not by inspection of syntax.  The
base case is $\sbt(\cdots, \num\,\mathbf{O}) = \sbt(\cdots,\mathbf{O}) =
\mathbf{O} = \num\,\mathbf{O}$; the step uses $\num(\mathbf{s}\,X) =
\seq{\mathbf{k}_{\mathrm{tag\_ap_1}}, \seq{\mathbf{k}_{\mathrm{tag\_s}},
\num\,X}}$ (the lemma $\mathbf{num\_at\_S}$), which at \emph{each} inductive
step rewrites the numeral to a $\seq{\cdot}$-code the descent can enter,
discharging the recursive call by the induction hypothesis.  So the descent
never meets a bare $\num(\mathbf{x}_1)$: it meets a planted num-leaf and
passes through it.  Lifting through the $\mathbf{ap_1}/\mathbf{ap_2}$
code-wrappers (the predicate $\mathit{NumCode}$ in \verb|BRA4.SbStep|),
\emph{every} substituent built from numeral leaves is inert under
\emph{every} single-variable substitution.

\paragraph{Intuition: the internalisation of ``$\mathbf{S}^n\mathbf{0}$ is
closed''.}  The lemma says nothing more than the elementary fact that a
numeral is a \emph{closed} term: $\mathbf{S}^n\mathbf{0}$ contains no
variable, so every substitution acts as the identity on it.  For a concrete
numeral this is immediate --- its code is a var-free $\code{\cdot}$-tree with
no $\seq{\mathbf{k}_{\mathrm{tag\_var}}, \cdot}$ leaf, so the encoded re-scan
visibly passes through.  The only subtlety, and the reason an internal
induction is needed at all, is the \emph{generic} statement: for a variable
$X$ the term $\num\,X$ is a \emph{stuck open} application, not literally a
numeral, so its closedness is invisible to the tag-dispatch and must be
\emph{promoted to an internal theorem}.  $\mathbf{ruleIndNat}$ is precisely
that promotion: it internalises ``$\num\,X$ is (the code of) a closed
numeral, for all $X$'' as a derivable equation.

\paragraph{Consequence: the multi-variable machinery is eliminated.}  With
inertness in hand, each compound-functor congruence of Theorem~12 --- e.g.\
specialising the binary congruence at the \emph{open} terms
$g(\underline{\mathbf{x}}), g(\mathbf{x}), h(\underline{\mathbf{x}}),
h(\mathbf{x})$ in building $D_{\mathbf{C}(f,g,h)}$ --- and the Step~5a wrap of
Theorem~14 are realised by \emph{nested single-$\mathbf{sb}$ wraps}: an inner
pass plants a substituent, an outer pass plants the next and re-scans the
inner one, which inertness leaves fixed.  Concretely $\thmT$ decodes a
$k$-fold nested $\seq{\mathbf{k}_{\mathrm{tag\_sb}}, \cdots}$ wrap by $k$
applications of its single $\mathbf{sb}$-clause into
$\sbf(\mathrm{spec}_0,\sbf(\mathrm{spec}_1,\cdots))$.  The dedicated two- and
three-variable functors $\sbttwo,\sbftwo,\mathbf{sbt_3},\mathbf{sbf_3}$, their
contract records, and the two extra discriminator levels in the $\thmT$
cascade --- roughly ten thousand lines across a dozen files --- are thereby
\emph{removed}.  This recovers exactly Guard's own explicit Theorem~14
construction (guard15.pdf p.~17), where the multi-variable effect is obtained
by \emph{nested single-$\mathbf{sb}$ steps with the open substituent applied
last}; what Guard left implicit --- that the re-scanned substituents
genuinely pass through --- is the inertness lemma.

\paragraph{The induction-avoiding alternative, and an open problem.}  There
is a second, equally sound route that we deliberately keep on record.  A
genuine \emph{simultaneous} substitution functor $\subtwo/\mathbf{sub_3}$
plops all substituents at once; nothing is re-scanned, so its closure proofs
need \emph{no} internal induction.  (It is sound: its encoded closures are
the encoded-level counterpart of the \emph{derived admissible} rule
$\mathbf{ruleInst2}$, $k_1\neq k_2 \Rightarrow {\vdash}\mathbf{P} \Rightarrow
{\vdash}\mathit{simSubstF}(k_1,t_1,k_2,t_2,\mathbf{P})$ --- Church's Standard
Metatheorem~VII, \emph{fully derived} from single substitution by
$\alpha$-renaming through a fresh variable in \verb|BRA3.RuleInst2| --- so it
adds no proof-theoretic strength.)  The contrast is precise: the
\emph{abstract} rule reduces to single substitution on raw formulas, but the
\emph{encoded} simultaneous functor does not reduce to iterated encoded
$\sbt$ \emph{by computation} (the re-scan stalls) --- only \emph{by
derivability}, and that derivation is exactly $\mathbf{sbt\_num\_inert}$,
which uses internal induction.  So the two routes differ in \emph{where they
spend induction}: the inertness route puts an internal induction into the
verifier's substitution-correctness; the simultaneous-substitution route
keeps that piece induction-free.

This bookkeeping is not idle.  Guard (guard19.pdf, p.~19, Def.~14 and the
\textsc{open problem} following Theorem~18) defines the \emph{induction-free}
fragment $\mathrm{BRA}_0$ --- the variable-free instances of BRA's axioms
with \emph{modus ponens as the only rule} --- whose verifier $\mathbf{th}_0$
has \emph{only} an axiom clause and an mp clause (no substitution clause, no
induction clause), and \emph{conjectures} that its consistency
$\mathbf{th}_0(y) \neq \code{\mathbf{O}=\mathbf{s}\,\mathbf{O}}$, though
valid, is \emph{unprovable in BRA} --- a putative counterexample to the usual
provable consistency of induction-free fragments.  Whether, and where, a
verifier's substitution machinery can be made induction-free is exactly the
kind of distinction that question turns on; the inert-lemma versus
simultaneous-substitution choice above is one concrete instance, which is why
we keep the simultaneous functors on record even though the shipped
development uses the single-variable route.

\section{Conclusion}

The formalisation closes Guard's argument for G\"odel's second
incompleteness theorem for BRA in a completely concrete,
constructively-verified form, and clarifies the implicit mathematical
machinery (closedness side conditions, predicate-Leibniz, Carneiro lift,
validating-decoder invariant, the two-forms-of-$\mathbf{G}$ distinction) on
which Guard's text relies without comment.


\end{document}